\documentstyle[12pt]{article}
\newcounter{mycount}

\newcommand{\be}{\begin{eqnarray}}
\newcommand{\ee}{\end{eqnarray}}
\newcommand{\bfl}{\begin{flushleft}}
\newcommand{\efl}{\end{flushleft}}

\newcommand\noi{\noindent}

\begin{document}

\bibliographystyle{nphys}

\centerline {FORM CONNECTIONS}
\vspace* {-25 mm}
\begin{flushleft} ITP 93-9 \\
May 1993
\end{flushleft}

\vskip 0.9in
\centerline{Ingemar Bengtsson }

\vskip 2cm
\centerline{\bf Abstract}
\vskip 2 cm

Riemannian geometry in four dimensions naturally leads to an SL(3) connection
that annihilates a basis for self-dual two-forms. Einstein's equations may
be written in terms of an SO(3) connection, with SO(3) chosen as an appropriate
subgroup of SL(3). We show how a set of "neighbours" of Einstein's equations
arises because the subgroup may be chosen in different ways. An explicit
example
of a non-Einstein metric obtained in this way is given. Some remarks on three
dimensional space-times are made.

\vskip 3mm
\vfill\noi
\begin {flushright} Institute of Theoretical  Physics

S-41296 G{\"o}teborg

Sweden
\end {flushright}

\vskip 3mm \noi

\eject

As is by now well known, it is possible to formulate Einstein's equations using
a configuration space that is spanned by an SO(3) connection \cite{Ash}
\cite{CDJ}.
 In a related development, 'tHooft \cite{tHo} found an action for the same
equations which employs an SL(3) connection as the basic variable. The purpose
of this letter is to use a part of the latter formalism to illuminate some
aspects of the former, in particular the surprising possibility \cite{Cap}
to construct diffeomorphism invariant theories which do not lead to Einstein
geometries, even though they have the same number of degrees of freedom as
Einstein's theory. Please note that many of the ideas which we will use have
been around for a long time \cite{Ple}.

  Although we will later make some comments on three-dimensional spaces, most
of
 our results are peculiar to four dimensions, and they all hinge on the fact
that
the six-dimensional space ${\bf W}$ of two-forms on a four-dimensional vector
space may be split into two orthogonal subspaces, with a scalar product given
by

\vspace{3mm}

\begin{equation}({\Sigma}_1,{\Sigma}_2) \equiv \frac{1}{2}
{\epsilon}^{{\alpha}{\beta}{\gamma}{\delta}}
{\Sigma}_{{\alpha}{\beta}1}{\Sigma}_{{\gamma}{\delta}2}.\end{equation}

\vspace{3mm}
\noindent
Moreover it turns out that there is a one-to-one correspondence
between all the ways in which ${\bf W}$ may be split into two orthogonal
subspaces ${\bf W^+}$ and ${\bf W^-}$ and the space of conformal structures
on the original vector space. Assuming that the triplet of two-forms
${\Sigma}_{{\alpha}{\beta}i}$ forms a basis for ${\bf W^+}$, it may be
shown that these basis vectors are by construction self-dual with respect to
the metric

\vspace{3mm}

\begin{equation}
g_{{\alpha}{\beta}} \equiv \frac{8}{3} {\sigma}
{\epsilon}^{ijk}{\Sigma}_{{\alpha}{\gamma}i}
\tilde{{\Sigma}}^{{\gamma}{\delta}}_{j}{\Sigma}_{{\delta}{\beta}k}
\end{equation}
\vspace{3mm}
\noindent
where the conformal factor ${{\sigma}}$ is so far arbitrary, the volume element
becomes

\vspace{3mm}
\begin{equation} \sqrt{-g} = \frac{i}{2} {\sigma}^{2} detm, \end{equation}

\vspace{3mm}
\noindent
and we use the notation

\vspace{3mm}

\begin{equation} \tilde{{\Sigma}}^{{\alpha}{\beta}} \equiv \frac{1}{2}
{\epsilon}^{{\alpha}{\beta}{\gamma}{\delta}}{\Sigma}_{{\gamma}{\delta}}\end{equation}

\vspace{3mm}
\begin{equation} m_{ij} \equiv {\epsilon}^{{\alpha}{\beta}{\gamma}{\delta}}
{\Sigma}_{{\alpha}{\beta}i}{\Sigma}_{{\gamma}{\delta}j}.\end{equation}

\vspace{3mm}
\noindent
The proof of this statement \cite{Urb} is an elegant application of Clifford
algebra ideas, and starts from the elementary observation that

\vspace{3mm}

\begin{equation}
{\Sigma}_{{\alpha}{\gamma}i}\tilde{{\Sigma}}^{{\gamma}{\beta}}_{j}
 + {\Sigma}_{{\alpha}{\gamma}j}\tilde{{\Sigma}}^{{\gamma}{\beta}}_{i} =
 - 1/4 m_{ij}{\delta}_{{\alpha}}^{{\beta}}.\end{equation}

\vspace{3mm}
\noindent
(In eq. (6) only, the index i is not restricted to run from one to three. This
 should cause no confusion.)

Note that we have chosen conventions appropriate to the case of Lorentzian
signatures, although it should be said that the conditions that guarantee that
we obtain a real metric with a particular signature is so far fully
understood (in the context of the field equations that we will present) only in
the
Euclidean and neutral cases.

With these facts, and the assumption that the ${\Sigma}'s$ are non-degenerate
in the sense that they may serve as a basis for the self-dual subspace, a pair
 of very useful formulas follows:

\vspace{3mm}

\begin{equation}
{\Sigma}_{\alpha}{}^{\gamma}{}_{i}{\Sigma}_{\gamma}{}^{\beta}{}_{j}
 = \frac{i}{8\sqrt{-g}} m_{ij} {\delta}_{{\alpha}}^{{\beta}} -
\frac{1}{2} {\sigma}^{-1}
{\epsilon}_{ijk}m^{km}{\Sigma}_{\alpha}{}^{\beta}{}_{m}\end{equation}

\vspace{3mm}

\begin{equation} {\Sigma}_{{\alpha}{\beta}i}m^{ij}{\Sigma}_{{\gamma}{\delta}j}
=
\frac{1}{8} {\epsilon}_{{\alpha}{\beta}{\gamma}{\delta}} -
\frac{i}{8\sqrt{-g}}(g_{{\alpha}{\gamma}}g_{{\beta}{\delta}} -
g_{{\alpha}{\delta}}g_{{\beta}{\gamma}})\end{equation}

\vspace{3mm}
\noindent
where

\vspace{3mm}

\begin{equation} m_{ik}m^{kj} \equiv {\delta}_{i}{}^{j} \end{equation}

\vspace{3mm}
\noindent
and Greek indices are raised and lowered with the metric (2).

  We observe that eq. (2) is invariant under SL(3) transformations acting
on theindex i. It is natural to try to define a connection that "annihilates"
the
two-forms ${\Sigma}_{{\alpha}{\beta}i}$, in the sense that

\vspace{3mm}

\begin{equation} \nabla_{\alpha}{\Sigma}_{{\beta}{\gamma}i} =
\partial_{\alpha}{\Sigma}_{{\beta}{\gamma}i} -
{\Gamma}_{{\alpha}{\beta}}{}^{{\delta}}{\Sigma}_{{\delta}{\gamma}i} -
{\Gamma}_{{\alpha}{\gamma}}{}^{{\delta}}{\Sigma}_{{\beta}{\delta}i} +
{\cal A}_{{\alpha}i}{}^{j}{\Sigma}_{{\beta}{\gamma}j},\end{equation}

\vspace{3mm}
\noindent
where ${\Gamma}_{{\alpha}{\beta}}{}^{{\gamma}}$ is a symmetric affine
connection.
 Counting components, we find that this will admit a solution for the
connection
whenever the latter is an SL(3) connection \cite{tHo} (SL(3) is
eight-dimensional),
and it may be verified that the affine connection then becomes metric
compatible,
provided that the conformal factor ${\sigma}$ is some function of the
${\Sigma}'s$.
 It is now straightforward to relate the SL(3) curvature to the Riemann tensor
of
the metric (2):

\vspace{3mm}

\begin{equation} 0 =
[\nabla_{\alpha},\nabla_{\beta}]{\Sigma}_{{\gamma}{\delta}i} =
 R_{{\alpha}{\beta}{\gamma}}{}^{{\sigma}}{\Sigma}_{{\sigma}{\delta}i} -
 R_{{\alpha}{\beta}{\delta}}{}^{{\sigma}}{\Sigma}_{{\sigma}{\gamma}i} +
 {\cal F}_{{\alpha}{\beta}i}{}^{j}{\Sigma}_{{\gamma}{\delta}j}.
\end{equation}

\vspace{3mm}
\noindent
Here ${\cal F}_{{\alpha}{\beta}}{}_{i}{}^{j}$ is the SL(3) curvature tensor.
It is easy to show that the symmetric part

\vspace{3mm}

\begin{equation} {\cal F}_{{\alpha}{\beta}(ij)} =
 {\cal F}_{{\alpha}{\beta}(i}{}^{k}m_{j)k} = 0 ,\end{equation}

\vspace{3mm}
\noindent
which means that the curvature lies in an SO(3) subgroup of SL(3).
 The connection may be written as

\vspace{3mm}

\begin{equation}  {\cal A}_{{\alpha}ij} \equiv {\cal A}_{{\alpha}i}{}^{k}m_{kj}
\equiv i{\epsilon}_{ikj}{\alpha}_{\alpha}{}^{k} +
{\beta}_{{\alpha}ij};\end{equation}

\vspace{3mm}

\begin{equation}{\beta}_{{\alpha}ij} = \frac{1}{2} (\partial_{\alpha} -
 {\Gamma}_{{\alpha}{\gamma}}{}^{{\gamma}})m_{ij}. \end{equation}

Eq. (11) may be solved for the SL(3) curvature, or alternatively for
the self-dual part of the Riemann tensor, with the result that

\vspace{3mm}

\begin{equation} R^{(+)}_{{\alpha}{\beta}{\gamma}{\delta}} =
\frac{{\sigma}}{2}{\epsilon}^{ijk}m_{km}
{\cal F}_{{\alpha}{\beta}j}{}^{m}{\Sigma}_{{\gamma}{\delta}i}\end{equation}

\vspace{3mm}

\begin{equation} {\cal F}_{{\alpha}{\beta}i}{}^{j} =
\frac{2i}{{\sigma}}\sqrt{-g}R_{{\alpha}{\beta}}{}^{{\gamma}{\delta}}
{\Sigma}_{{\gamma}{\delta}k}{\epsilon}_{imn}m^{mj}m^{nk}. \end{equation}

\vspace{3mm}

Now we turn to dynamics. In writing down field equations, we prefer to
use an SO(3) connection, since the formalism simplifies. Moreover SO(3)
admits spinor representations while SL(3) does not. So we switch to the
formalism of ref. \cite{CDJ}, and introduce an SO(3) connection, its
field strength, and the matrix

\vspace{3mm}

\begin{equation}{\Omega}^{ij} \equiv {\epsilon}^{{\alpha}{\beta}
{\gamma}{\delta}}F_{{\alpha}{\beta}}{}^{i}F_{{\gamma}{\delta}}{}^{j}.\end{equation}

\vspace{3mm}
\noindent
Then we write an action of the general form \cite{Cap}

\vspace{3mm}

\begin{equation} S = \int {\cal L}({\eta};Tr{\Omega}, Tr{\Omega}^{2},
Tr{\Omega}^{3})\end{equation}

\vspace{3mm}
\noindent
where ${\eta}$ is a Lagrange multiplier as well as a scalar density of
weight minus one, and the traces are defined by means of the group metric
of SO(3), e.g.

\vspace{3mm}

\begin{equation}Tr{\Omega} \equiv g_{ij}{\Omega}^{ij}
.\end{equation}

\vspace{3mm}
\noindent
In order to keep things reasonably specific, we will concentrate on the special
case

\vspace{3mm}

\begin{equation} S = \frac{1}{8} \int {\eta}(Tr{\Omega}^{2} +
 {\alpha}(Tr{\Omega})^{2}), \end{equation}

\vspace{3mm}
\noindent
where ${\alpha}$ is a parameter. It is shown in ref. \cite{CDJ}
that this action gives field equations that are equivalent to Einstein's
equations for vanishing cosmological constant, provided that a certain
non-degeneracy condition holds,

\vspace{3mm}

\begin{equation} {\alpha} = - \frac{1}{2} ,\end{equation}

\vspace{3mm}
\noindent
and provided that the metric is defined by

\vspace{3mm}
\begin{equation}g_{{\alpha}{\beta}} = \frac{8}{3}{\eta}{\epsilon}_{ijk}
F_{{\alpha}{\gamma}}{}^{i}\tilde{F}^{{\gamma}{\delta}j}F_{{\delta}{\beta}}{}^{k} .
\end{equation}

\vspace{3mm}

What about the general case? It is very important to observe that, an
arbitrary conformal factor apart, the definition of the metric is not
arbitrary.
It is enforced by the requirement that the theory should have a Hamiltonian
formulation which makes sense. This requirement also permits one to show
that the same formula for the metric holds (up to conformal transformations)
 in the general case, so the previous equation is in fact a theorem.
(For details, see ref. \cite{Cap}.) However, this metric is not Einstein
unless ${\alpha} = - 1/2$. The Hamiltonian formulation moreover shows that
the number of degrees of freedom remains two per space-time point for all
the theories that can be constructed in this way, which is why we refer to
them as "neighbours of Einstein's equations". We observe that, by construction,

\vspace{3mm}

\begin{equation} F_{{\alpha}{\beta}i} \in {\bf W^+} \end{equation}

\vspace{3mm}
\noindent
where ${\bf W^+}$ denotes the self-dual subspace.

Let us consider the field equations from the action (20):

\vspace{3mm}

\begin{equation} {\epsilon}^{{\alpha}{\beta}{\gamma}{\delta}}
D_{{\beta}}{\Sigma}_{{\gamma}{\delta}i} = 0 \end{equation}

\vspace{3mm}

\begin{equation} {\Sigma}_{{\alpha}{\beta}i} \equiv
 {\Psi}_{ij}F_{{\alpha}{\beta}}{}^{j} ;
\hspace{4mm} {\Psi}_{ij} \equiv 2{\eta}({\Omega}_{ij} +
 {\alpha}g_{ij}Tr{\Omega}) \end{equation}

\vspace{3mm}

\begin{equation} Tr{\Omega}^{2} + {\alpha}(Tr{\Omega})^{2} = 0 . \end{equation}

\vspace{3mm}
\noindent
In the second equation (which from now on is used to define ${\Sigma}$) we
have,
for the first and last time in this letter, used the group metric to lower
indices.
 One can show that the choice of conformal factor in eq. (22) implies that

\vspace{3mm}

\begin{equation} \sqrt{-g}g_{{\alpha}{\beta}} =
\frac{i}{6} {\epsilon}^{ijk}({\Sigma}_{i}\tilde{{\Sigma}}_{j}
{\Sigma}_{k})_{{\alpha}{\beta}}. \end{equation}

\vspace{3mm}
\noindent
Hence the conformal factor of the metric can be expressed as a function
of the ${\Sigma}'s$.

The first field equation may be used to solve for the connection in terms
 of the ${\Sigma}'s$. The calculation is straightforward and the result is
important, but its explicit form is unilluminating, so we do not give it here.
Of more immediate interest is the following consequence of the third equation
($m_{ij}$ is defined in eq. (5)):

\vspace{3mm}

\begin{equation} m_{ij} =
4{\eta}^{2}\left ((1 + 2{\alpha}){\Omega}^{2}{}_{ij}Tr{\Omega} +
 ({\alpha} + \frac{1}{2})({\alpha} - 1){\Omega}_{ij}(Tr{\Omega})^{2}\right ) -
8i\sqrt{-g}g_{ij}. \end{equation}

\vspace{3mm}
\noindent
(Here $\sqrt{-g}$ denotes the square root of the space-time metric, not the
group metric.) Hence the two natural metrics that exist on the SO(3) fibers
are proportional if and only if ${\alpha} = - 1/2$, that is to say in the
 Einstein case.
How are the SO(3) and SL(3) formalisms related? Introducing a metric on the
fibers breaks SL(3) down to SO(3). However, in general we have two different
fibre metrics available, $m_{ij}$ and $g_{ij}$. Eq. (28) reveals that they
are proportional only if ${\alpha} = - 1/2$, that is to say in the Einstein
case. In general we are dealing with two different SO(3) subgroups. The
SO(3) field strength is self-dual by construction, and we see that

\vspace{3mm}

\begin{equation} {\cal F}_{{\alpha}{\beta}i}{}^{j} \in {\bf W^+} \hspace{4mm}
\Leftrightarrow \hspace{4mm} m_{ij} \propto g_{ij} .\end{equation}

\vspace{3mm}
\noindent
Referring back to eq. (15), we see that the self-dual "from the right"
part of the Riemann tensor is self-dual "from the left" - a condition which
implies the vanishing of the traceless part of the Ricci tensor - only when
 the two fibre metrics are proportional. We also see from eq. (28) that
this is a quite exceptional case.

To add concreteness to the above, we will now consider a simple exact
solution of eqs. (24-26). We choose the Ansatz \cite{CP}

\vspace{3mm}

\begin{equation} A_{x1} = a_{1}(t) \hspace{4mm} A_{y2} = a_{2}(t)
\hspace{4mm} A_{z3} = a_{3}(t), \end{equation}

\vspace{3mm}
\noindent
all others zero ($A_{{\alpha}i}$ is the SO(3) connection). If we choose the
gauge

\vspace{3mm}

\begin{equation} {\eta} = \frac{1}{16\dot{a}_1\dot{a}_2\dot{a}_3}
\end{equation}
\vspace{3mm}
\noindent
we find the solution

\vspace{3mm}

\begin{equation} a_{i} = t^{- \frac{2c_{i}}{c}} \hspace{3mm} ;
\hspace{5mm} c \equiv c_{1} + c_{2} + c_{3}, \end{equation}

\vspace{3mm}
\noindent
where the integration constants have to obey

\vspace{3mm}

\begin{equation} c_1{}^2 + c_2{}^2 + c_3{}^2 + {\alpha}c^2 = 0. \end{equation}

\vspace{3mm}
\noindent
The resulting metric is

\vspace{3mm}

\begin{equation} ds^2 = - dt^2 + t^{2{\gamma}_1}dx^2 + t^{2{\gamma}_2}dy^2 +
 t^{2{\gamma}_3}dz^2. \end{equation}

\vspace{3mm}
\noindent
The Kasner exponents are defined by

\vspace{3mm}

\begin{equation} {\gamma}_i \equiv 1 - \frac{2c_i}{c} \hspace{3mm} ;
\hspace{6mm} \sum {\gamma}_i = 1 \hspace{5mm} \sum {\gamma}_i{}^2 =
 - 1 - 4{\alpha}. \end{equation}

\vspace{3mm}
\noindent
For ${\alpha} = - 1/2$ this is the familiar Kasner metric. For other
values of ${\alpha}$ the Ricci tensor turns out to have only one
non-vanishing component, viz.

\vspace{3mm}

\begin{equation} R_{tt} = 2(1 + 2{\alpha})t^{-2}. \end{equation}

\vspace{3mm}

Finally, let us comment on the situation when the space-time dimension
differs from four. It is unlikely that geometry can be based on two-forms
in any dimension higher than four. In three dimensions, on the other hand,
there is a natural isomorphism between the space of two-forms and the space
 of vectors, which we express explicitly through

\vspace{3mm}

\begin{equation} *{\Sigma}^{\alpha} \equiv \frac{1}{2}
 {\epsilon}^{{\alpha}{\beta}{\gamma}}{\Sigma}_{{\beta}{\gamma}} .\end{equation}

\vspace{3mm}
\noindent
Therefore, in a three dimensional space-time a "form connection" as
defined by eq. (10) becomes just the ordinary SO(1,2) spin connection.
 There is now no analogue of the four dimensional "neighbours" of Einstein's
 equations, since there is essentially only one object available from which
 the action may be constructed, namely det*F. Actually, this is not quite true,
 since a Chern-Simons term may be added. Consider

\vspace{3mm}

\begin{equation} S = 2 \int \sqrt{ \frac{1}{{\lambda}} det*F} +
{\theta} {\epsilon}^{{\alpha}{\beta}{\gamma}}(A_{{\alpha}I}
\partial _{\beta}A_{\alpha}^I +
\frac{1}{3}f_{IJK}A_{\alpha}^IA_{\beta}^JA_{\gamma}^K).\end{equation}

\vspace{3mm}
\noindent
The first term here gives Einstein's equations with ${\lambda} =
 $ the cosmological constant \cite{Pel}. Including the second
term gives a non-vanishing torsion tensor

\vspace{3mm}

\begin{equation} T_{{\alpha}{\beta}}{}^I =
2{\theta}F_{{\alpha}{\beta}}{}^I . \end{equation}

\vspace{3mm}
\noindent
However, the field strength may still be expressed in terms of a
"metric" triad, so that the equation for the Ricci tensor eventually
collapses to

\vspace{3mm}

\begin{equation} R_{{\alpha}{\beta}} =
2{\lambda}(1 - {\lambda}{\theta}^2)g_{{\alpha}{\beta}}. \end{equation}

\vspace{3mm}
\noindent
We see that the Chern-Simons term only rescales the cosmological constant.
Its presence will be felt when matter fields are added.

\vspace{3mm}
\noindent
{\it Acknowledgements:} This work was done while I was visiting
Imperial College, London. I am grateful for the hospitality,
and to the Science Research Council for financial support.

\end{document}